\documentclass[12pt, a4paper]{article}

\begin{document}

\title{Comment on the hadronic decay of excited heavy quarkonia}

\author{Mu-Lin Yan\\ 
International Center for Theoretical Physics,\\
P.O.Box 586, 34100, Trieste, Italy\\
and\\
Center for Fundamental Physics\\
University of Science and Technology of China\\
Hefei, Anhui 230026, P.R.China\thanks{Mailing Address}\\
and\\
Yi Wei, Ting-Liang Zhuang \\
Center for Fundamental Physics and Center for Nonlinear Science \\
University of Science and Technology of China\\ 
Hefei, Anhui 230026, P.R.China }
\maketitle
\begin{center} {\bf To appear in The European Physical Journal}\end{center}
\begin{abstract}
We make comments on ref.\cite{me}, and provide partial wave analysis to 
the decays of excited heavy S-wave $1^-$ quarkonia into the basic $1^-$ 
quarkonia state plus $\pi\pi$. 
It is revealed that there exist contributions of
D-wave transition in $\psi'\longrightarrow J/\psi\pi\pi$,
$\Upsilon(2S)\longrightarrow \Upsilon(1S) \pi \pi$ and
$\Upsilon(3S)\longrightarrow \Upsilon(1S) \pi \pi$ by using the 
data-fitting results in ref.\cite{me}. A possible experimental method 
to measure the D-wave directly is discussed.
\end{abstract}

\paragraph{1, Introduction}: Starting from the infinite mass limit for the heavy quarkonium,  
the authors of \cite{me} have presented an interesting
systematic derivative expansion
for the decays of a heavy excited S-wave spin-1 quarkonium into
a lower S-wave spin-1 state in the spirit of chiral perturbation theory.
An effective Lagrangian for these exclusive hadronic decays has been
constructed in \cite{me}. It is of as follows
\begin{eqnarray}
{\cal L} & = & {\cal L}_{0} + {\cal L}_{\rm SB} \\
{\cal L}_{0} &= & gA_{\mu}^{(v)} B^{(v) \mu*}{\rm Tr}[(\partial_{\nu}U)
    (\partial^{\nu}U)^{\dagger}]\nonumber \\
& &    +g_1A_{\mu}^{(v)} B_{\nu}^{(v)*
    }{\rm Tr}[(v{\cdot}{\partial} U)(v{\cdot}{\partial} U)^{\dagger}]\nonumber
    \\
&&    +g_2 A_{\mu}^{(v)}B^{(v) \mu *}{\rm Tr} [({\partial}^{\mu}U)
    ({\partial}^{\nu}U)^{\dagger}+({\partial}^{\mu}U)^{\dagger}({\partial}^{\nu}
    U)]+h.c.   \\
{\cal L}_{\rm SB} &=& g_3 A^{(v)}_{\mu} B^{(v){\mu}*} {\rm Tr}
    [{ M}(U+U^{\dagger}-2)] \nonumber\\
&&    +ig^{'} {\varepsilon}^{\mu\nu\alpha\beta}[v_{\mu}A^{({v})}_{\nu}
    {\partial}_{\alpha}B^{(v)*}_{\beta}-{\partial}_{\mu}A^{(v)}
    _{\nu}v_{\alpha}B^{({v})*}_{\beta}]{\rm Tr}[{ M}
    (U-U^{\dagger})] \nonumber \\
&&    +h.c.
\end{eqnarray}
    where $ U$ is a unitary 3$ {\times}$3 matrix that contains the Goldstone
    fields, ${ M}$ is light-quark mass matrix, 
and $g$, $g_1$, $g_2$
, $g_3$ and $g'$ are constants. And $A_\mu$ is the field
of initial S-wave $1^-$ quarkonium state, $v$ is its velocity vector,
$B_\mu$ the one of the final
$1^-$ quarkonium state, and
    $A^{({v})}_\mu$
    and $B^{({v})}_\mu$ are defined by a phase redefinition of $A_\mu$ and
    $B_\mu$ respectively.
By using this Lagrangian, the authors of ref.\cite{me}
have obtained a good fit to the data,
especially to $\psi'\longrightarrow J/\psi \pi\pi$, 
$\Upsilon(2S)\longrightarrow \Upsilon(1S)\pi\pi$
and $\Upsilon(3S)\longrightarrow \Upsilon(1S)\pi\pi$. 
However, they do not present a correct
partial wave analysis to the decay amplitudes. The decays are S-wave 
transition dominant, but in principle D-wave transition is not forbidden,
and to search the signals of D-wave in the transitions is meaningful.
We will point out in this present comment on \cite{me} that to view the
$g_1$-term in ${\cal L}_0$ (eq.(2)) to be a S-wave transition purely is
a misunderstanding. We will show that the D-wave transition 
signals can be seen in the processes of
$\psi'\longrightarrow J/\psi \pi\pi$, 
$\Upsilon(2S)\longrightarrow \Upsilon(1S)\pi\pi$
and $\Upsilon(3S)
\longrightarrow \Upsilon(1S)\pi\pi$. And the ratios 
between D-wave contributions to the transitions and the total are determined.
In addition, we will also propose a possible experiment to measure 
the D-wave transitions directly, and will discuss how to determine all
constans in the process of $A_\mu\longrightarrow B_\mu\pi\pi$ experimently
besides a overall factor.

\paragraph{2, Kinematics and Partial wave analysis:} 
For definiteness, we consider the case of $A_\mu={\psi'}$ and $B_\mu=J/\psi$
in this section.
In the rest frame of $\psi^\prime$, $p_{\pi^+}$ and $p_{\pi^-}$ 
are 4-momentum of
$\pi^+$ and $\pi^-$ respectively, $q=p_{\pi^+}+p_{\pi^-}$, 
$r=p_{\pi^+}-p_{\pi^-}$ and the partial decay
rate of $\psi^\prime$ into $J/\psi \pi^+ \pi^-$ is given by
\begin{eqnarray}
d\Gamma &=& A_{\rm ph} \left | \cal M \right |^2dm_{\pi\pi}d\cos \theta
\end{eqnarray}
where ${\cal M}$ is the decay amplitude,
\begin{eqnarray*}
A_{\rm ph}&=&{1 \over {2 \pi}^2}{1 \over {16m^2_{J / \psi}}}
(q^2-4m^2_{\pi})^{1 \over 2} \left | \stackrel{\rightarrow} {q} \right |  \\
\left | \stackrel{\rightarrow} {q} \right | &=&{1 \over 2m_{\psi^\prime}}
[(m^2_{\psi^\prime}-(m_{\pi\pi}+m_{J/\psi})^2)
(m^2_{\psi^\prime}-(m_{\pi\pi}-m_{J/\psi})^2)]^{1 \over 2}   \\
q^2 &=& m^2_{\pi\pi}=(p_{\pi^+}+p_{\pi^-})^2,
\end{eqnarray*}
and $\theta$ is the angle between 3-momentum of $\pi^+$ and 3-momentum of
 $J/{\psi}$  in the rest frame of $\pi^+-\pi^- $. We call $\theta $ as 
correlation angle hereafter.
From the Lagrangian of eq.(1), one can get the transition amplitude 
${\cal M}$ to the process of $A_\mu=\psi'$ and $B_\mu=J/\psi$ as follows
\begin{eqnarray}
&&{\cal M} ( {\psi}^{'} {\rightarrow} J/{\psi}{\pi}^{+}{\pi}^{-}) \nonumber \\
& =& -{\frac {4}{F^{2}_{0}}} \{
  [ {\frac{g}{2}}(m^2_{\pi\pi}-2 M^2_{\pi})
 +g_1 (v {\cdot} p_{{\pi} ^{+}} )( v {\cdot} p_{{\pi} ^{-}})
+g_3 M_{\pi} ^{2} ] {\varepsilon}^{*}_{\psi}
 {\cdot}{\varepsilon}_{{\psi} ^{'}} \nonumber \\
&& +g_{2} [p_{{\pi} ^{+} {\mu}} p_{{\pi} ^{-} {\nu}} +p_{{\pi}^{+}{\nu}}
 p_{{\pi}^{-}{\mu}}]{\varepsilon}^{*{\mu}}_{J/{\psi}}{\varepsilon}^
 {\nu}_{{\psi}^{'}}
  \}
\end{eqnarray}
 where 
 $\varepsilon _{J/{\psi}}, \varepsilon _{\psi ^{'}} $ are the polarization
 vectors of $J/\psi$ and $\psi'$ respectively.

The authors of ref.\cite{me} argued that the $g_2-$terms in ${\cal M}$
(eq.(5)) are strongly suppressed by the chiral symmetry breaking scale
$\Lambda_{\chi SB}$ or heavy quark mass according to their derivative
expansion treatment, and they then set $g_2=0$. This argument is consistent
with the results achieved in \cite{vo},\cite{sh} and\cite{yan} based on
the multipole expansion hypothesis for the soft gluon field emission from
heavy quarkonium.
However this does not
mean the D-wave contribution has been excluded. And to view
this as pion-D-wave contribution suppression is inadequate,
because the $g_1-$term has D-wave component also. We show this point 
in follows.
                                                 
In the rest frame of $\psi^\prime$, $v=(1, 0)$
\begin{eqnarray*}
(v \cdot p_{\pi^+} )(v \cdot p_{\pi^-})
 =p_{\pi^+}^0p_{\pi^-}^0
\end{eqnarray*}
Through elementary calculations we get
\begin{eqnarray*}
p_{\pi^+}^0p_{\pi^-}^0
&=& A(q^2)P_0(\cos \theta)+B(q^2)P_2(\cos \theta) \\
\end{eqnarray*}
where
\begin{eqnarray*}
A(q^2)&=&\frac{1}{4}q^2+ \frac{1}{6}
\left |\stackrel\rightarrow {q} \right |^2
(1+\frac{2m_{\pi}^2}{q^2}) \\
B(q^2)&=&-\frac{1}{2} \left |\stackrel\rightarrow {q} \right |^2
(1-\frac{4m^2_{\pi}}{q^2})
\end{eqnarray*}
$ P_0(\cos \theta)=1 $,
$ P_2(\cos \theta)=\frac{1}{2}(\cos^2\theta-\frac{1}{3}) $
are Legendre functions.
Thus the decay amplitude $\cal M$ (eq. (5)) can be decomposed into two parts:
S-wave($\cal M_{\rm S}$) and D-wave( $\cal M_{\rm D}$).
We have
\begin{eqnarray}
   \cal M =\cal M_{\rm S}+\cal M_{\rm D}
\end{eqnarray}
where
          \begin{eqnarray}
{\cal M}_{\rm S}&=&
{\cal M}_{0} \{ q^{2}-c_1(q^{2}+
  \left | \stackrel{\rightarrow} {q} \right |
^2) (1+{{2m}^2_{\pi} \over {q^2}}) +c_2 m^2_{\pi} \}\\
{\cal M}_{\rm D} &=& {\cal M}_0
 \{ 3c_1
 \left | \stackrel {\rightarrow} q \right |^2
 (1-\frac{4m_{\pi}^2}{q^2}) \}P_2(\cos \theta)
\end{eqnarray}
with
\begin{eqnarray*}
  {\cal M}_0 &=& {\rm {const.}} \times(\varepsilon_{\psi^\prime}\cdot
\varepsilon_{J/\psi})  \\
c_{1} &=& -\frac{g_1}{3g}(1+\frac{g_1}{6g})^{-1}   \\
c_2 &=& 2(\frac{g_3}{g}-\frac{g_1}{3g}-1)(1+\frac{g_1}{6g})^{-1}
\end{eqnarray*}
The ratio of D-wave transition rate to the total decay rate is defined by
\begin{eqnarray}
{\cal R}_{\rm D} &=& \frac{
  {\int dq^2 \int^{+1}_{-1} {d\cos  \theta}
{\frac{1}{m_{\pi\pi}}}
A_{\rm ph} \sum\limits_{\varepsilon \varepsilon'} 
\left | \cal M_{\rm D} \right |^2}  }{
{\int dq^2 \int^{+1}_{-1} d\cos  {\theta}{\frac{1}{m_{\pi\pi}}}
A_{\rm ph} \sum\limits_{\varepsilon \varepsilon'}
\left | {\cal M}_{\rm S}+{\cal M}_{\rm D} \right  | ^2} }
\end{eqnarray}
where the limits of $q^2$ in the integrals are $q^2_{\rm min}=4m^2_{\pi}$
,$q^2_{\rm max}=( M_{\psi ^\prime}-M_{J/\psi})^2$
and the data used
 in the calculations are $ M_{\psi^{'}}=3686.0  MeV$, 
 $M_{J/ \psi}= 3096.88  MeV$,
   $ M_{\pi}=139.57  MeV$ and $\sum\limits_{\varepsilon \varepsilon'}$ means 
to sum both up $\varepsilon_{J/\psi}$ and up $\varepsilon_{\psi'}$. 
Thus as long as
 $g_1$ and $g_3$ be fixed by fitting the experimental invariant mass
 spectrum, the contributions of D-wave to the transitions will be determined.

The extensions of the above formulas to the excited $\Upsilon$-decays are
straightforward.

\paragraph{3, Ratios of D-wave transition rate to the total decay rate:}
From eqs.(4) and (5), we obtain the invariant $\pi-\pi$-mass spectrum
\begin{eqnarray}
\frac{{d\Gamma}}{ d {m_{\pi\pi}}}&=&
\int_{-1}^{1} d\cos \theta A_{ph}\sum\limits_{\varepsilon \varepsilon'}
|{\cal M}|^2  \nonumber \\
&=&\Gamma_0 \left | \stackrel {\rightarrow} q\right | \sqrt{q^2-4m^2_{\pi}}
\{ [q^2-c_{1}(q^2+ \left | \stackrel {\rightarrow} q\right |^2)(1+
\frac{2m_{\pi}^2}{q^2})+c_{2} m_{\pi}^2]^2
\nonumber \\
&& +{1\over 5}{c_{1}}^2
\left | \stackrel {\rightarrow}q\right |^4(1-\frac {4m_{\pi}^2}{q^2})^2 \}.
\end{eqnarray}
where $\Gamma_0$ is a constant. 

To $\psi'\longrightarrow J/\psi\pi\pi$
and $\Upsilon'\longrightarrow \Upsilon\pi\pi$, the energies that are 
available for the pions are small ($<586 MeV<m_\rho$). Therefore the fit
under chiral limit, i.e., $g_3=0$, to these processes is legitimate.
This has been done in ref.\cite{me}. The authors of \cite{me} obtained
\begin{eqnarray}
({g_1\over g})_{c\bar{c}}^{\rm chiral}&=&-0.35\pm0.03, \;\;\;
{\rm for}\;\; \psi'\longrightarrow J/\psi \pi^+ \pi^-, \\
({g_1\over g})_{b\bar{b}}^{\rm chiral}&=&-0.19\pm0.04, \;\;\;
{\rm for}\;\; \Upsilon(2S)\longrightarrow \Upsilon(1S) \pi^+ \pi^-. 
\end{eqnarray}
Substituting eq.(11) and eq.(12) into eq.(9), we obtain the ratios of
D-wave transition rate to the total rate for 
$ \psi'\longrightarrow J/\psi \pi^+ \pi^-$ and
$ \Upsilon(2S)\longrightarrow \Upsilon(1S) \pi^+ \pi^- $ respectively
as follows 
\begin{eqnarray}
{\cal R}_D(\psi'\longrightarrow \psi \pi\pi)&=&0.065\pm0.018\% \\
{\cal R}_D(\Upsilon (2S)\longrightarrow \Upsilon (1S) \pi\pi)
&=&0.0156\pm0.0078\% .
\end{eqnarray}
To $\Upsilon (3S)\longrightarrow \Upsilon (1S) \pi\pi$, 
the energies that are available for the pions are not small, and the pions
are not soft. Thus the low energy theorem based on the chiral symmetry
may be not a good approximation. Consequently, the contributions of $g_3$-term
turned to be significant.
$g$, $g_1$ and $g_3$ of this process have been determined\cite{me} to be
$$ {g_1\over g}=-2.86\pm0.37,\;\;\; {g_3\over g}=15.0\pm1.2. $$
(These values should be checked by a further fit of the correlation angle
spectrum. To see the next section.)
Then the D-wave content for this process is determined
\begin{equation}
{\cal R}_D(\Upsilon (3S)\longrightarrow \Upsilon (1S) \pi\pi)
=35.5\pm14.2\% .
\end{equation}

To $\psi' \longrightarrow \psi\pi\pi $, in order to reveal the contributions 
of $g_3-$term of eq.(1) ( an effect due to chiral symmetry breaking), the
authors have designed a fitting procedure to fit the experimental 
invariant $\pi-\pi$ mass spectrum. However, it is of a lack of ground in
physics. We will show in the next section that the value of $g_3$ can
be determined by fitting both the invariant mass spectrum and the 
correlation angle spectrum of this process.

\paragraph{4, Direct measurement to D-wave transition:}
The fact that there exist a small amount of D-wave transitions in the
processes of $\psi' \longrightarrow J/\psi\pi\pi$, $\Upsilon'\longrightarrow
\Upsilon\pi\pi$ etc indicates that these decays are not exactly isotropic
to $\theta$-distributions. To $\psi' \longrightarrow J/\psi\pi\pi$,
the $\theta$-distributions can be explored by the correlation angle
spectrum of the process, which is as follows
\begin{eqnarray}
\frac{\rm {d\Gamma}}{\rm d{\cos \theta}}
&=&\int_{2m_\pi}^{M_{\psi'}-M_{J/\psi}} dm_{\pi\pi} A_{ph} |{\cal M}|^2
 \nonumber \\
&=&\rm const.{\Gamma_0}(1+0.18c_{2}+0.0085{c_{2}}^2
-3.54c_{1}-0.35 c_{1}c_{2}+3.54{c_{1}}^2  \nonumber  \\
&&+(0.77c_{1}+0.077c_{1}c_{2}-1.62c_{1}^2)
(\cos^2\theta-\frac{1}{3})    \nonumber  \\
&& +0.20 c_{1}^2 (\cos^2\theta -\frac{1}{3})^2)
\end{eqnarray}
A fit to this spectrum represents a direct measurement to the partial waves
in the decay.

In the decay amplitude ${\cal M}$ without $g_2$-terms (eq.(6)), there are
two unknown parameters $g_1/g$ and $g_3/g$. They could be determined by
fitting two measured curves: the invariant mass spectrum (eq.(10)) and
the correlation angle ($\theta$) spectrum (eq.(16)). The fit of the former
has been performed in \cite{me}, and fit of the latter is expected. Because 
the ratios of D-wave transition rate to the total are generally less than
$1\%$, it is not easy to see the deviations of 
$\theta$-distribution from the isotropic $\theta$-spectrum. However,
it is essential that a great quantity of $\psi'$ ($\sim 4\times 10^6$)
have been accumulated by Beijing Electron Spectrum (BES) on BEPC\cite{BES}.
This will make the measured curve of $\theta$-spectrum accurate enough
to exhibit the the D-wave transition effects in the decay.
To $\Upsilon (3S)\longrightarrow\Upsilon(1S)\pi\pi$, it is necessary to
fit this $\theta$-spectrum in order to check the corresponding result of
\cite{me}.

Finally, we like to mention that so long as the number of the events are 
large enough, it is practicable and helpful too to fit the measured 
bi-variable spectrum of follows
\begin{equation}
{d^2\Gamma \over dm_{\pi\pi}d\cos\theta}
=\sum\limits_{\varepsilon\varepsilon'} A_{ph}|{\cal M}|^2.
\end{equation}
This is a two-dimensional fit.
Because both $m_{\pi\pi}$ and $\theta$ are not integrated out,
more interesting informations on the dynamics of the decays are left in
this bi-variable spectrum. A full expression of ${\cal M}$ is eq.(5)
where there are three parameters, $g_1/g,\; g_2/g$ and $g_3/g$. We like
to argue that a full fit to the spectrums of eq.(10), eq.(16) and eq.(17)
will provide useful informations for these parameters. Since $g_2$-terms
in ${\cal M}$(eq.5) describe the physics beyond its leading order effects,
the informations on it would be significant to the dynamics.

\paragraph{5, Summary:} 
An interesting systematic derivative expansion for the decays of a heavy 
excited S-wave spin-1 quarkonium into a lower S-wave spin-1 state is
presented in\cite{me}. In this letter, we make some comments on its results.
We provide a partial wave analysis to the decays of excited heavy 
S-wave $1^-$ quarkonia into the basic $1^-$ quarkonia plus $\pi\pi$.
It is revealed that there exist contributions of
D-wave transition in $\psi'\longrightarrow J/\psi\pi\pi$,
$\Upsilon(2S)\longrightarrow \Upsilon(1S) \pi \pi$ and
$\Upsilon(3S)\longrightarrow \Upsilon(1S) \pi \pi$ by using the 
data-fitting results in ref.\cite{me}. A possible experimental method 
to measure the D-wave directly is discussed.
It is expected to measure the process of $\psi'\longrightarrow\psi\pi\pi$
more precisely by using the data accumulated by BES on BEPC.
We argue that through measuring the invariant mass spectrum, the correlation
angle spectrum and the bi-variable spectrum precisely, three parameters in
the model of \cite{me} could all be determined.

\begin{center} {\bf Acknowledgments} \end{center}
We would like to thank J Li and Z J Guo (BES, Beijing) for helpful discussions.
One of us (MLY) wishes to acknoledge the International Center for Theoretical
Phsics, Trieste, for its hospitality where part of this work was done. This work was
supported in part by the National Funds of China throgh C N Yang, and the Funds of
the Institute for High Energy Physics, Beijing.

\end{document}